\documentclass[prl,aps,twocolumn,footinbib,superscriptaddress,floatfix,bibliography]{revtex4-1}

\usepackage{CJK}
\usepackage[colorlinks,bookmarks=false,citecolor=blue,linkcolor=red,urlcolor=blue]{hyperref}
\usepackage{color} 
\usepackage{graphicx}
\usepackage{float}
\usepackage{multirow}
\usepackage{amsmath}
\usepackage{bm}
\usepackage{amsfonts}
\usepackage{braket}
\usepackage{subfigure}
\usepackage{cleveref}

\begin{document}

\title{Topological Route to New and Unusual Coulomb Spin Liquids}

\author{Owen Benton}
\affiliation{Max Planck Institute for the Physics of Complex Systems, N{\"o}thnitzer Str. 38, Dresden 01187, Germany}

\author{Roderich Moessner}
\affiliation{Max Planck Institute for the Physics of Complex Systems, N{\"o}thnitzer Str. 38, Dresden 01187, Germany}

\begin{abstract}
Coulomb spin liquids are topological magnetic states obeying an emergent Gauss’ law. Little distinction has been made between different kinds of Coulomb liquids.
Here we show how a series of {\it distinct} Coulomb liquids 
can be generated straightforwardly by varying the constraints on a classical spin system. This leads to pair creation/annihilation, and coalescence, of topological defects of an underlying vector field. The latter makes higher-rank spin liquids, of recent interest in the context of fracton theories, with attendant multi-fold pinch points in the structure factor, appear naturally. New Coulomb liquids with an abundance of pinch points also arise. 
We thus establish a new and general route to uncovering exotic Coulomb liquids, via the manipulation of topological defects in momentum space.
\end{abstract}

\maketitle

One of the most appealing 
aspects of many-body systems
is their ability to produce large scale behavior 
described by emergent degrees of freedom very different from the system's microscopic constituents. The emergence of a gauge field at low energy  is taken as a defining feature of a topological state of matter. 
A prominent example occurs in the Coulomb phase of spin ice \cite{Harris1997, Bramwell2001, Isakov2004,Isakov2005, Hermele2004,
Fennell2009, Castelnovo2012}, in which the strongly interacting spins are described by the fluctuations of a (otherwise free) field constrained by Gauss' law:
\begin{eqnarray}
\nabla \cdot {\bf B}=0.
\end{eqnarray}

Analogous descriptions exist for
other frustrated  models \cite{Moessner1998,Henley2005,Henley2010,Rehn2016} in which the low energy configurations obey a constraint which can be encoded as a Gauss' law for an emergent field or fields. The resulting state is called 
a Coulomb spin liquid, in analogy to the Coulomb phase of a U(1) gauge theory.

Recent work has highlighted extensions in which the emergent field is a higher-rank symmetric tensor $E_{\mu \nu}$, and obeys a 
generalisation of Gauss' law
\cite{Xu2006, Rasmussen2016, Pretko2017, Prem2018, Yan2020}, e.g.:
\begin{eqnarray}
\partial_{\mu} E_{\mu \nu}=0, \ \forall \ \nu.
\end{eqnarray}
Higher rank Coulomb liquids are of particular interest due to a close connection with fracton theories~\cite{Nandkishore2019}, sought after for 
their exotic nature  and their potential usefulness in quantum information contexts \cite{Schmitz2018}.

There are only few examples of simple spin models exhibiting Coulomb liquids, with higher-rank cases rarer still. Also, we lack a systematic understanding for how to distinguish, and account for, various different types of Coulomb liquid. Further, there has been little exploration of possible transitions between  types, or of how lower-rank Coulomb liquids relate to higher rank ones.

Here we introduce an approach which, besides filling in those gaps, enables the discovery, and provides a description of, new Coulomb liquids in classical spin systems almost mechanically, and shows how higher rank Coulomb liquids arise at transitions between distinct lower-rank ones.
In so doing, we establish a new understanding of the topological nature of these spin liquids.

\begin{figure}
    \centering
    \subfigure[]{\includegraphics[width=0.31\columnwidth]{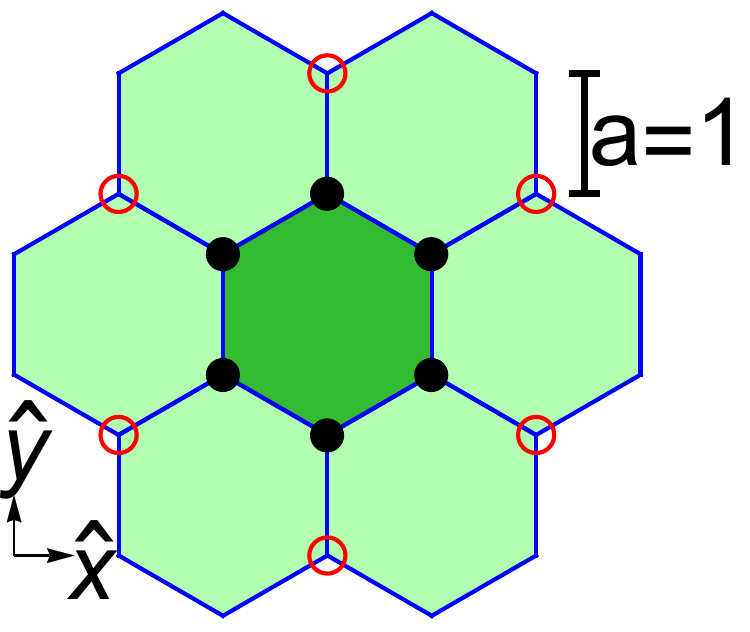}}
    \subfigure[]{\includegraphics[width=0.64\columnwidth]{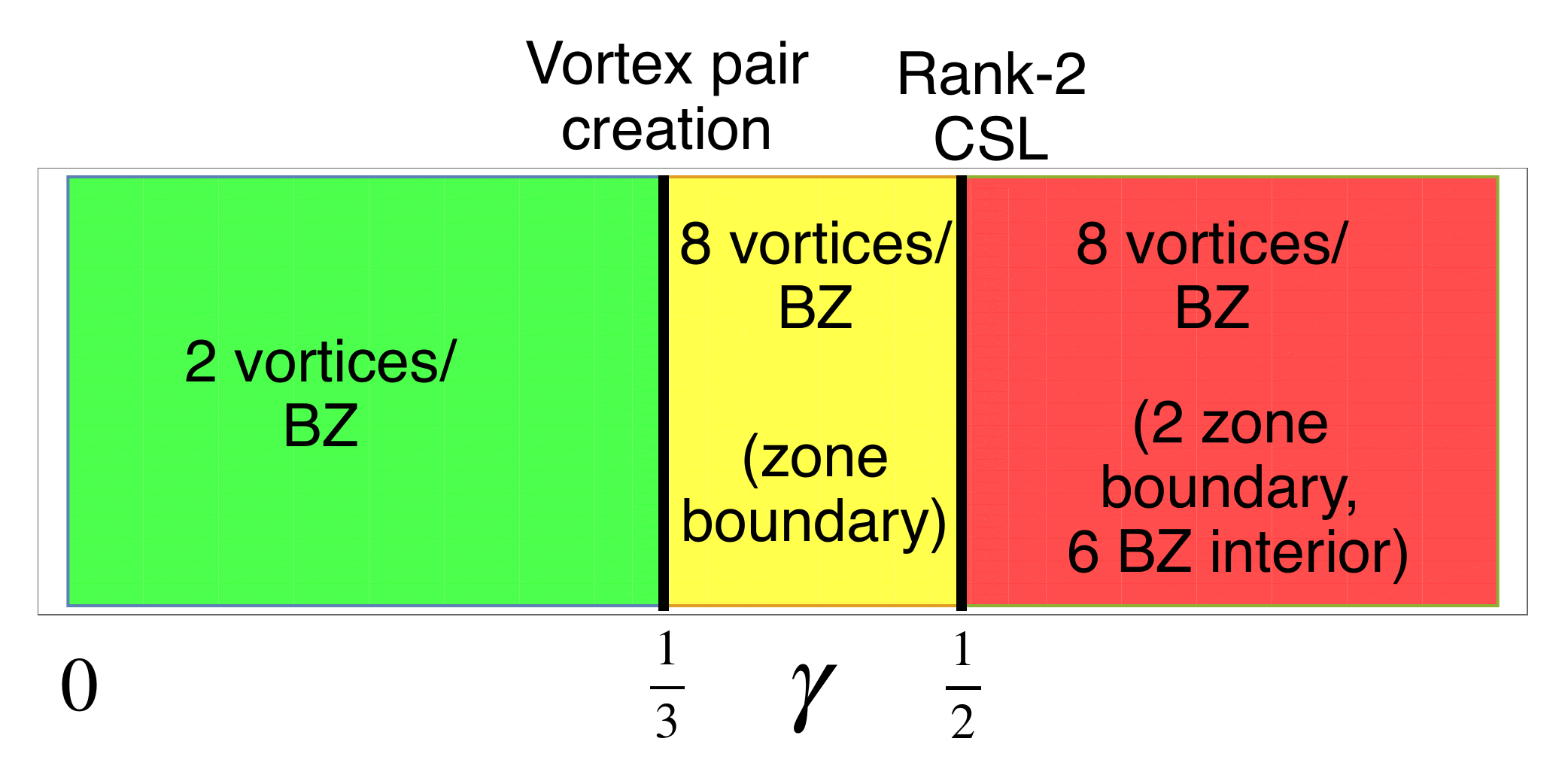}}
    \caption{Honeycomb lattice model exhibiting a series of distinct classical spin liquids. (a) The local constraint defining the ground states of the model. 
    The sum of spins on each hexagon (black, filled circles) plus a coefficiennt $\gamma$ multiplied by the sum of spins linked to the exterior of the hexagon (red, open circles) must vanish on every hexagon [Eqs. (\ref{eq:Mhex})-(\ref{eq:honeycomb_constraint})].
    (b) Phase diagram as a function of  $\gamma$, showing a series of  algebraically correlated CSLs. They are distinguished by the number and arrangement of topological defects of the constraint vector ${\bf L}({\bf q})$ in the Brillouin zone, with these defects giving rise to pinch points in the structure factor $S({\bf q})$ [Fig. \ref{fig:honeycomb_sq}]. Transitions between spin liquids occur either by creation/annihilation of pairs of defects ($\gamma=1/3$) or by coalescence of defects with a net charge ($\gamma=1/2$) leading to a higher-rank Coulomb liquid with multi-fold pinch points.}
    \label{fig:honeycomb}
\end{figure}

\begin{figure*}
    \centering
   \includegraphics[width=1.0\textwidth]{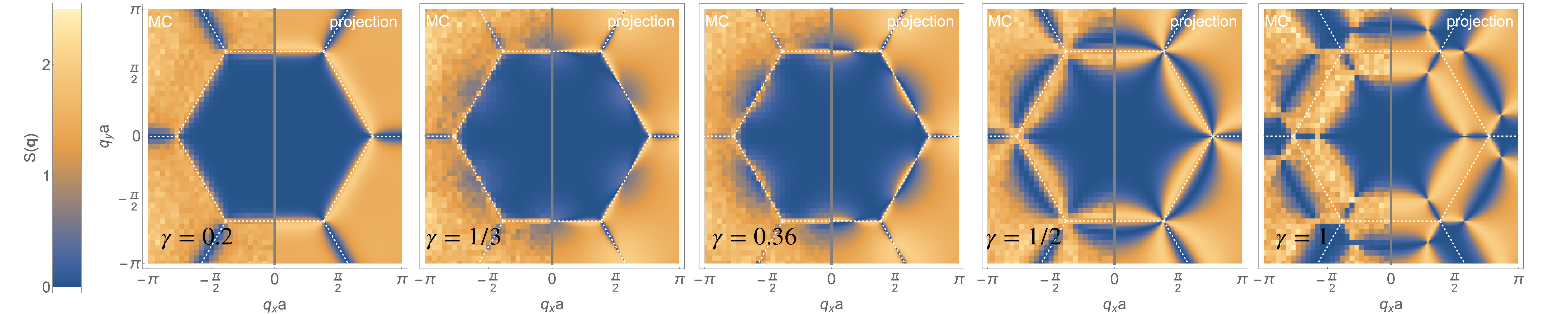}
    \caption{Evolution of $S({\bf q})$ across distinct honeycomb spin liquids [Fig. \ref{fig:honeycomb}]. For small values of $\gamma$ there are vortices in ${\bf L}({\bf q})$ and corresponding pinch points in $S({\bf q})$ at the Brillouin Zone corners. At $\gamma=1/3$ there is a nucleation of pairs of oppositely charged topological defects at the zone boundary which then migrate towards the zone corners.
    At $\gamma=1/2$ the coalescence of these defects into objects with charge $Q=\pm2$ manifests in the appearance of 4-fold pinch points in the structure factor, associated to the emergence of a Rank-2 Coulomb liquid. On increasing  $\gamma$ further these defects separate and the system enters a new spin liquid with 6 pinch points in the interior of the Brillouin zone, as well as at the zone corners.
    The left half of each panel is the result of a Monte Carlo simulation of $N=1920$ spins at $T=0.002J$ and the right half is a calculation using the projection approach [Eq. (\ref{eq:Sq_proj})].
    }
    \label{fig:honeycomb_sq}
\end{figure*}

\begin{figure}
    \centering
  \subfigure[ ]{\includegraphics[width=0.3\columnwidth]{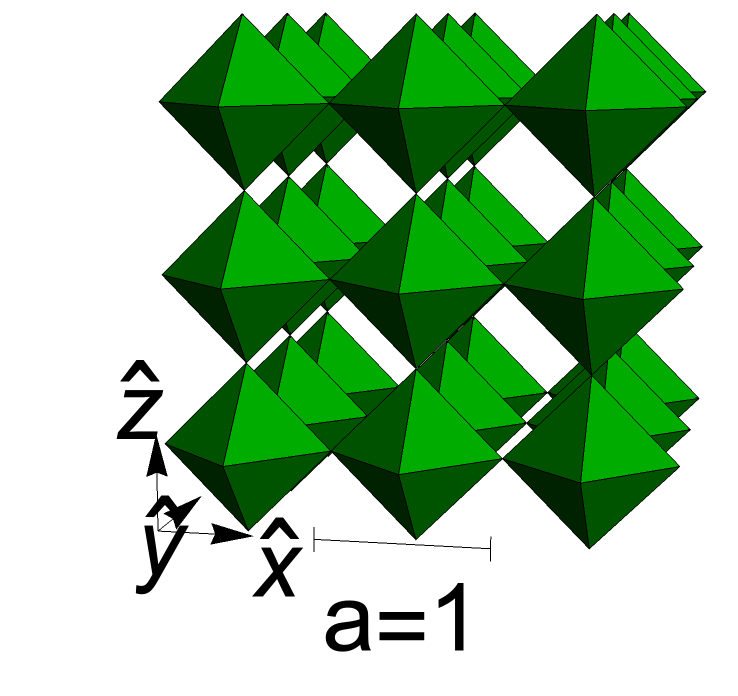}}
  \subfigure[ ]{\includegraphics[width=0.3\columnwidth]{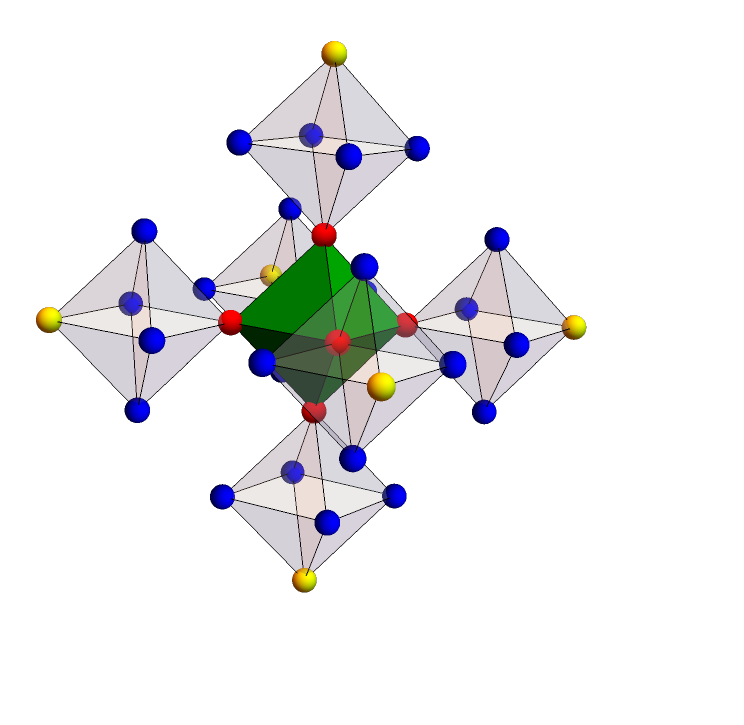}}
  \subfigure[ ]{\includegraphics[width=0.6\columnwidth]{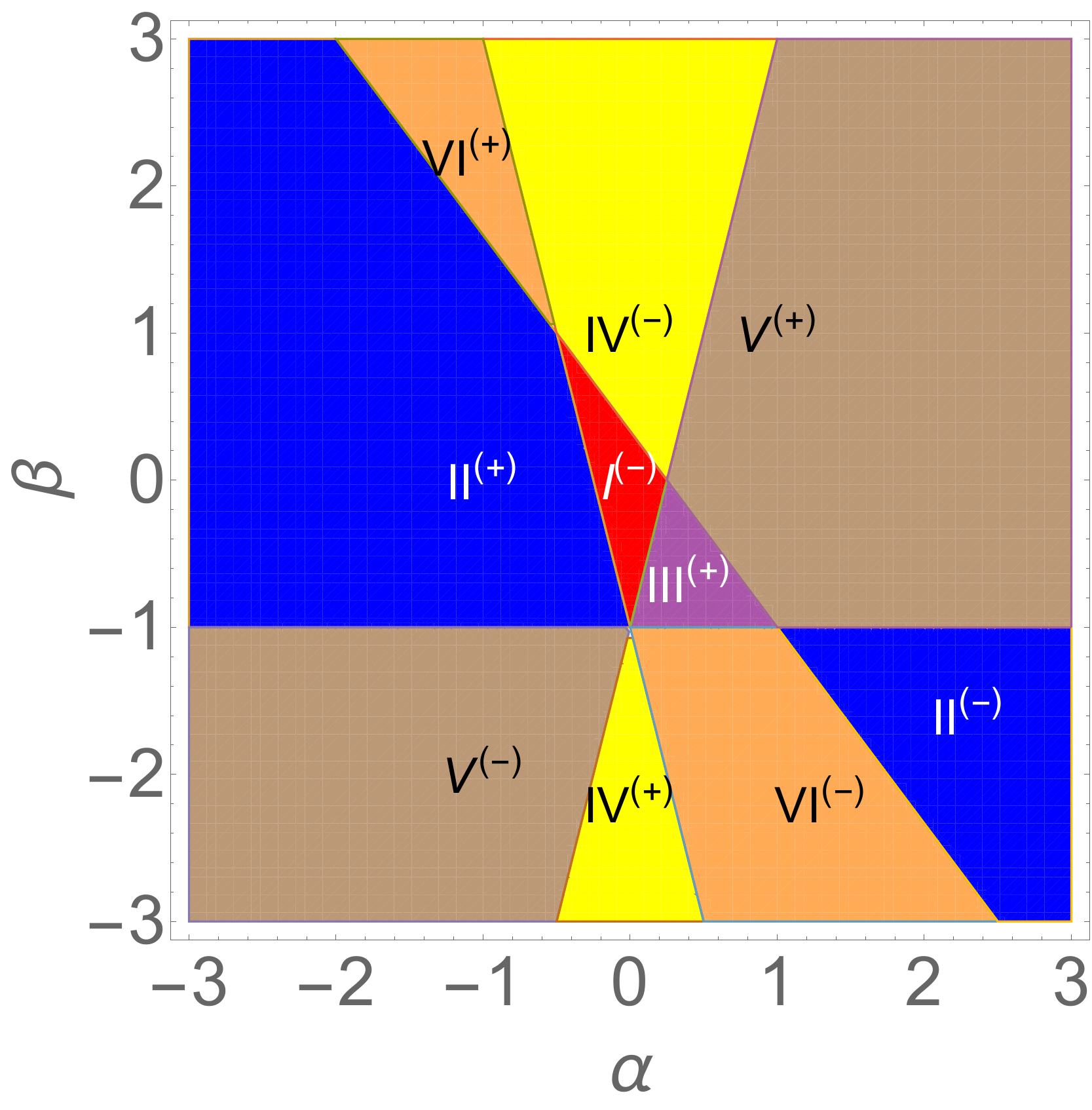}}
    \caption{Construction and phase diagram of octochlore spin liquids. (a) The octochlore lattice, a network of corner sharing octahedra, is the medial lattice of the simple cubic lattice. (b) The local constraint applying to each octahedron in the lattice. The sum of the spins on the central octahedron (red), plus $\alpha$ times the sum the spins  neighboring that octahedron (blue), plus $\beta$ times the sum of remaining spins on the surrounding octahedra (yellow) must vanish [Eq. (\ref{eq:Moct})].  (c) The phase diagram of distinct Coulomb spin liquids obtained by varying $\alpha$ and $\beta$, distinguished by the number and arrangement of topological defects in ${\bf L}({\bf q})$.
    For the definition of each spin liquid see Table \ref{tab:octo_table}.
    }
    \label{fig:octochlore}
\end{figure}

\begin{figure}
    \centering
  \subfigure[ ]{\includegraphics[width=0.43\columnwidth]{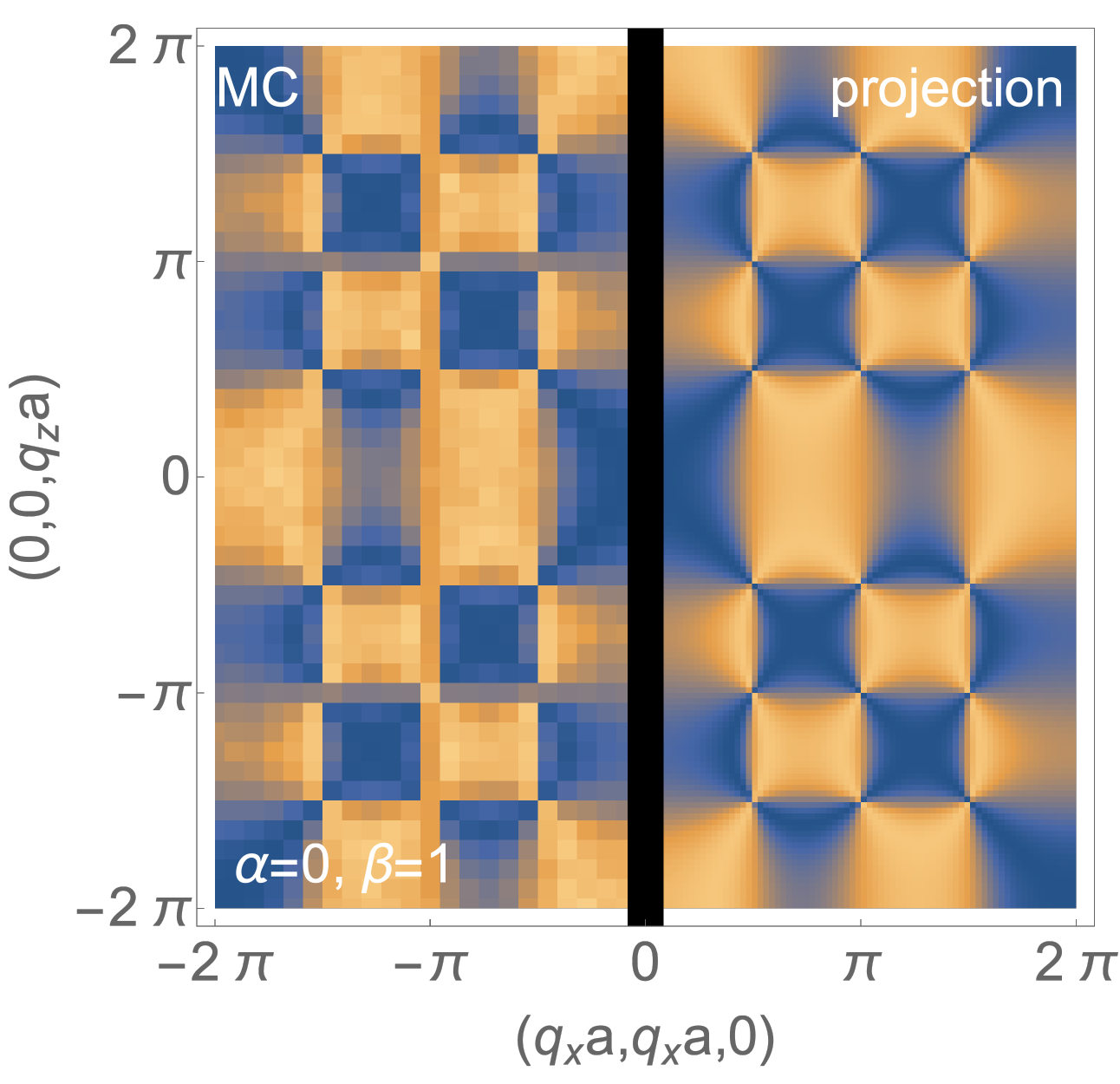}}
  \subfigure[ ]{\includegraphics[width=0.43\columnwidth]{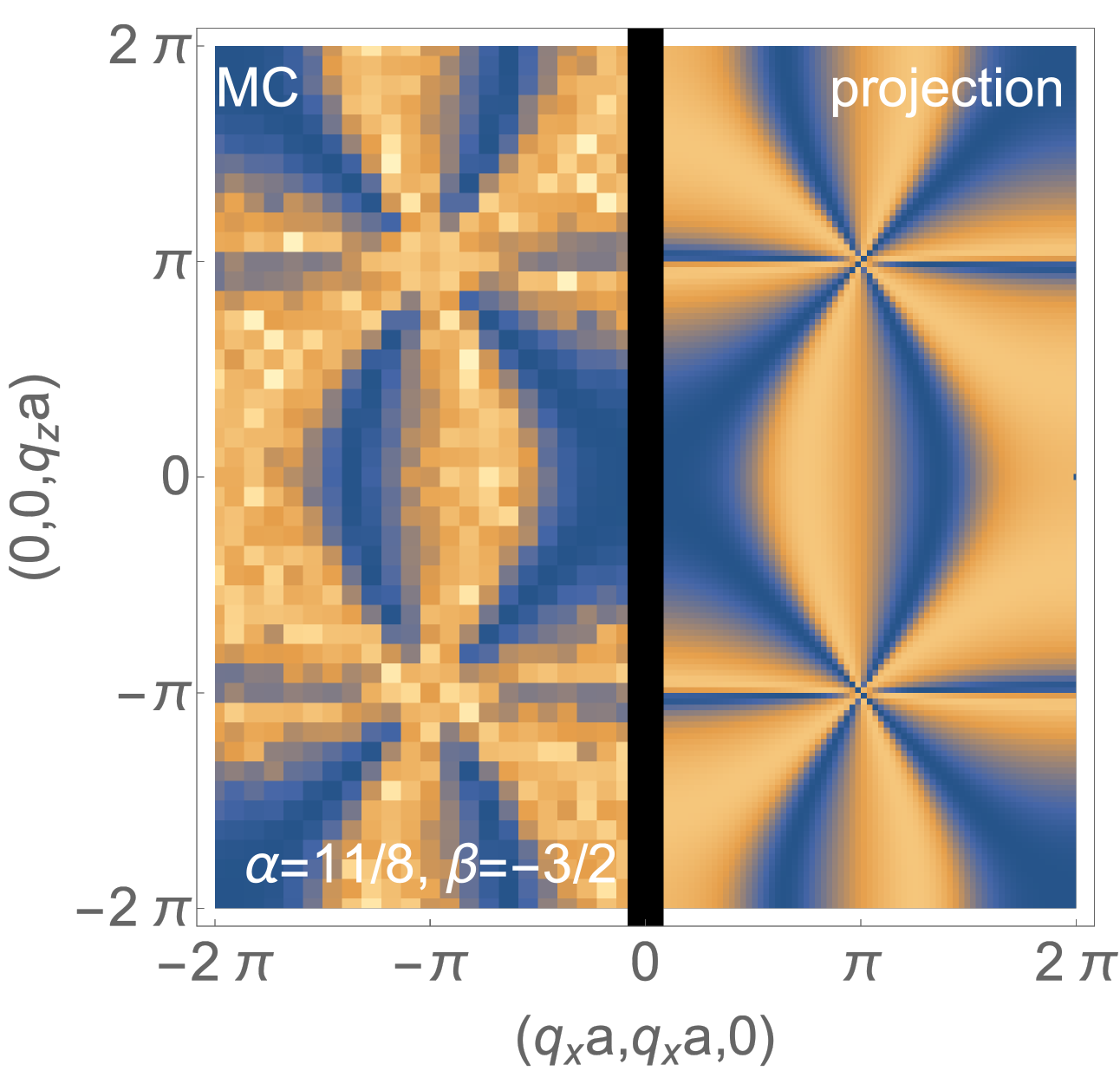}} \
  \includegraphics[width=0.09\columnwidth]{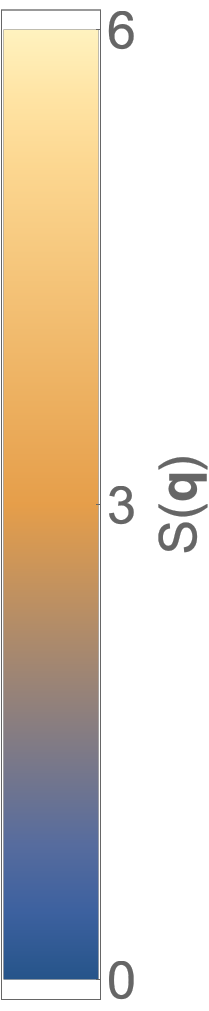}
    \caption{Abundant and multi-fold pinch points in octochlore spin liquids. 
    (a) $S({\bf q})$ in region IV of Fig.  \ref{fig:octochlore}(c). ${\bf L}({\bf q})$ exhibits topological defects at  $(\pi, \pi, \pi)$ and at wavevectors removed along the $(1,0,0)$,
    $(1,0,0)$ and $(1,1,1)$ directions from that point. The structure factor exhibits pinch point singularities at the corresponding positions.
    (b) $S({\bf q})$ at the boundary between regions II and VI of Fig.  \ref{fig:octochlore}(c).
    There are $Q=7$ topological defects at the zone corners which manifest as multi-fold pinch points, associated to a higher rank Coulomb liquid.
    The left half of each panel is the result of a Monte Carlo simulation of $N=31 944$ spins at $T=0.01J$ and the right half is a calculation using the projection approach.
    }
    \label{fig:octochlore_sq}
\end{figure}

In the remainder of this paper we first introduce the general framework, which we then illustrate using two classical spin models, one on the honeycomb lattice [Fig. \ref{fig:honeycomb}-\ref{fig:honeycomb_sq}] and
one on the octochlore lattice [Fig. \ref{fig:octochlore}-\ref{fig:octochlore_sq}] (corner sharing octahedra) which between them realise a multitude of new and unusual Coulomb liquids.

{\it General approach:}
Our approach is based on relating the ground state constraints in real space (Eq.~(\ref{eq:Mhex}) in the honeycomb example below) imposed by the spin liquid's Hamiltonian (Eq.~(\ref{eq:Hhex})) to the topological properties of a vector function in momentum space ${\bf L}({\bf q})$.
For Heisenberg spins, the real-space constraint quite generally takes the form
\begin{eqnarray}
\sum_{i \in c} \eta_i {\bf S}_i =0 \ \forall \ c
\label{eq:real_space_constraint}
\end{eqnarray}
where $c$ are some real space clusters (e.g. triangles in the kagome lattice, tetrahedra in the pyrochlore lattice) and $\eta_i$ are some real coefficients \footnote{Eq. (\ref{eq:real_space_constraint}) can be generalized to
cover anisotropic classical spin liquids \cite{Benton2016-pinchline, Taillefumier2017} by replacing $\eta_i\to\eta_{i\mu}$, ${\bf S}_i\to S_{i\mu}$ where $\mu$ is a spin component index and summing over both $\mu$ and $i$.}.

If the system has translational symmetry, then Eq. (\ref{eq:real_space_constraint}) can be rewritten in momentum space:
\begin{eqnarray}
&&\sum_{m=1}^{n_{u}}  L_m^{(p)} ({\bf q})^{\ast} {\bf S}_m({\bf q})   =0 \ \ \forall  \ \ {\bf q}, \ p 
\label{eq:q_space_constraint}
\\
&&L_m^{(p)}=\sum_{i \in m \in c_p} \eta_i \exp(i {\bf q} \cdot ({\bf r}_{c_p}-{\bf r}_i))
\label{eq:L_def}
\end{eqnarray}
where there are $n_u$ sites per unit cell, and the sum in Eq. (\ref{eq:q_space_constraint}) runs over the $n_u$ translationally inequivalent sublattices. ${\bf S}_m({\bf q})$ is the lattice Fourier transform of the spin configuration on the $m^{th}$ sublattice. Eq. (\ref{eq:L_def}) defines  a set of $n_u$-component vectors, with members of the set being indexed by $(p)$. The number of vectors ${\bf L}^{(p)}$ is equal to the number of real space constraints in one unit cell. The sum in Eq. (\ref{eq:L_def}) runs over spins belonging to sublattice $m$
within a single constrained cluster. ${\bf r}_{i}-{\bf r}_{c_p}$ is the position of spin $i$ relative to the center of the cluster.

This 
representation is inspired by Henley \cite{Henley2005}, who related spin correlation
functions to a projection matrix $\mathcal{P} ({\bf q})$,
enforcing Eq. (\ref{eq:q_space_constraint}) by projecting out all ${\bf L}^{(p)}({\bf q})$:
\begin{eqnarray}
\langle {\bf S}_m(-{\bf q}) \cdot 
{\bf S}_n({\bf q}) \rangle= \frac{1}{\kappa} 
\mathcal{P}_{mn} ({\bf q})
\label{eq:Sq_proj}
\end{eqnarray}
$m$ and $n$ are sublattice indices and $\kappa$ is a normalisation constant to enforce the sum rule on the correlation function \footnote{Eq. (\ref{eq:Sq_proj}) is an approximation to the spin correlation function, which by construction respects Eq. (\ref{eq:q_space_constraint}) exactly, but respects spin normalisation only on average.}.
Eq. (6) is an approximation: specifically it may be considered as a zero-temperature limit of the leading order of a $1/\mathcal{N}$ expansion, with $\mathcal{N}$ the number of spin components.
However, we emphasize that this approximation is not
necessary to define ${\bf L}^{(p)}({\bf q})$, although it is helpful in establishing the
relationship between ${\bf L}^{(p)}({\bf q})$ and the spin correlation
functions.

Points in ${\bf q}$-space where one of the ${\bf L}^{(p)}({\bf q})$ vanishes, or where one ${\bf L}^{(p)}({\bf q})$ becomes linearly dependent on the others, have special significance since at those momenta Eq. (\ref{eq:q_space_constraint}) is satisfied trivially for at least one value of $p$.
At such points the projection matrix $\mathcal{P}_{mn}({\bf q})$ becomes singular, and there will be a corresponding singularity in the structure factor (a pinch point in most known examples).

Crucially, depending on $n_u$, and the symmetries of the system, the zeros of ${\bf L}^{(p)}({\bf q})$ can carry a topological charge. They will therefore be robust against small modifications to the constraint (\ref{eq:real_space_constraint}), provided that the  symmetries necessary to define the charge are maintained. Larger modifications can cause pair-creation or annihilation of topological defects, and thereby a transition to a different spin liquid
with more/fewer singularities in the {\bf q}-space correlations.
By manipulating Eq. (\ref{eq:real_space_constraint}) one can
also coalesce topological defects into higher-charged objects.

We find that manipulations of this kind can generate new Coulombic spin liquids and  transitions between them. Further, the coalescence of ${\bf q}$ space defects into objects with higher topological charge is associated to the formation of multi-fold pinch points in the structure factor, (i.e. pinch points with more than 2 intense lobes). Such singularities are signatures of higher-rank spin liquids described by tensor gauge theories \cite{Prem2018, Yan2020}, and we therefore establish a simple mechanism for the construction of higher-rank Coulomb spin liquids.

\emph{Honeycomb Model:} We consider a model on the honeycomb
lattice [Fig. \ref{fig:honeycomb}], with the Hamiltonian:
\begin{eqnarray}
&&\mathcal{H}_{\sf h}=
\frac{J}{2} \sum_{{\rm hex}} {\bf M}_{{\rm hex}, \gamma}^2\label{eq:Hhex}
\\
&&{\bf M}_{{\rm hex}, \gamma}=\sum_{i \in {\rm hex}} {\bf S}_i + \gamma \sum_{i \ \in \langle {\rm hex} \rangle} {\bf S}_i 
\label{eq:Mhex}.
\end{eqnarray}
The sum in Eq. (\ref{eq:Hhex}) is over hexagonal plaquettes of the lattice.
The first sum in  Eq. (\ref{eq:Mhex}) is over spins on the plaquette (filled black circles in Fig. \ref{fig:honeycomb}(a)) and the second is over spins connected to the outside of the plaquette (red open circles in Fig. \ref{fig:honeycomb}(a)). 
$\gamma$ is a dimensionless parameter with which we tune the model.
Ground states of Eq. (\ref{eq:Hhex}) everywhere satisfy the constraint 
\begin{eqnarray}
{\bf M}_{{\rm hex},\gamma}=0 \ .
\label{eq:honeycomb_constraint}
\end{eqnarray}
For $\gamma=0$ this model reduces to the one studied in \cite{Rehn2016}.

We this constraint in Fourier space, using
Eqs. (\ref{eq:q_space_constraint})-(\ref{eq:L_def}).
With only one constraint per unit cell, i.e.\ only one vector ${\bf L} ({\bf q})$, we can drop the index $(p)$.
There are two sites per unit cell, so ${\bf L} ({\bf q})$ has two, complex, components: $L_{1}({\bf q}),
L_{2}({\bf q})$, which can be calculated from Eq. (\ref{eq:L_def}).
Since Eq. (\ref{eq:q_space_constraint}) is invariant under
rescaling ${\bf L} ({\bf q})$, we normalize $\tilde{{\bf L}} ({\bf q}) = {{\bf L}} ({\bf q})/|{{\bf L}} ({\bf q})|$.

Inversion symmetry imposes that $\tilde{L}_{1}({\bf q})=\tilde{L}_{2}({\bf q})^{\ast}$.
$\tilde{{\bf L}} ({\bf q})$ can thus be written as
\begin{eqnarray}
\tilde{{\bf L}} ({\bf q})=
\frac{1}{\sqrt{2}}( \exp(i \phi_{{\bf q}}), \exp(-i \phi_{{\bf q}})).
\end{eqnarray}
Thus, $\tilde{{\bf L}} ({\bf q})$ for the honeycomb model [Eq. (\ref{eq:Hhex})] can always be mapped to a position on the
unit circle, and can host stable vortices. 
At the position of these vortices ${\bf L}({\bf q})$ must vanish, and there is a corresponding singularity in the spin correlation function [Eq. (\ref{eq:Sq_proj})].

The vortices have an integer winding number defined for 
closed paths C in momentum space:
\begin{eqnarray}
Q_C=\frac{-1}{\pi} \oint_C d{\bf q}\cdot\left(
\tilde{L}_2 \nabla_{\bf q} \tilde{L}_1
\right).
\end{eqnarray}
The total winding number is conserved under smooth
changes to the constraint Eq. (\ref{eq:real_space_constraint}) (e.g. varying $\gamma$). New 
vortices can only be created in oppositely charged pairs.

At $\gamma=0$, calculation of $\tilde{{\bf L}}({\bf q})$ 
reveals vortices at the corners of the
Brillouin Zone (BZ) and their positions coincide as expected
with the location of pinch point singularities in the structure factor \cite{Rehn2016}
\begin{eqnarray}
S({\bf q})=\sum_{mn} \langle {\bf S}_m(-{\bf q}) \cdot 
{\bf S}_n({\bf q}) \rangle.
\label{eq:Sq_def}
\end{eqnarray}

The evolution of $S({\bf q})$ at finite $\gamma$ is shown in
Fig. \ref{fig:honeycomb_sq}.
The left half of each panel shows $S({\bf q})$ calculated from a Monte
Carlo simulation \cite{supplemental}, 
the right half a calculation using the projection method
[Eq. (\ref{eq:Sq_proj})] \cite{Henley2005}.
The good agreement between the two indicates that
conclusions drawn from analysis of ${\bf L}({\bf q})$,
the basis of the projection calculation,
are robust.

For  $\gamma<1/3$ the only vortices are those at the
BZ corners, and the corresponding pinch points in $S({\bf q})$ remain robust.
At $\gamma=1/3$ pairs of vortices
with opposite winding numbers nucleate  at the zone boundaries.
For $\gamma>1/3$ these vortex pairs separate along the zone boundary.
There are now 8 rather than 2 vortices per BZ,
and the same number of pinch points in S({\bf q}).
The appearance of these new singularities in the structure
factor demonstrates that the system is in a qualitatively
distinct Coulomb liquid [Fig. \ref{fig:honeycomb}(b)].

As $\gamma$ increases further, the vortices migrate along the zone 
boundary. At $\gamma=1/2$, three positive (negative) vortices converge on a single negative (positive) vortex at the zone corner, to make vortices with winding number $Q=\pm2$.
This has a striking consequence for S({\bf q}): the appearance of four-fold pinch point singularities.
These are known as signatures of higher rank Coulomb
spin liquids with tensor electromagnetic fields \cite{Prem2018, Yan2020}.
Indeed, on coarse graining the model we find that
the spin liquid at $\gamma=1/2$ can be described in terms
of fluctuations of a traceless, symmetric tensor $m_{\mu \nu}$ subject
to the constraint:
\begin{eqnarray}
\partial_{\mu} \partial_{\nu} m_{\mu \nu}=0.
\label{eq:r2-m}
\end{eqnarray}
The relationship between
$m_{\mu \nu}$ and the microscopic spins is given in the Supplemental Material \cite{supplemental}.
The components of $m_{\mu \nu}$ are formed from the local order parameter for antiferromagnetic order 
with wavevector at the BZ corners.
Assuming slow variation of $m_{\mu \nu}$ in real space
(or, equivalently, expanding the ${\bf q}$ space constraints around the BZ corners), one can use a gradient expansion to turn the microscopic contraints
on the spin configuration into constraints on the spatial variation of $m_{\mu \nu}$.
At $\gamma=1/2$ the first derivative term in this expansion vanishes, leaving a second derivative term given by Eq. (\ref{eq:r2-m}), which is a generalized Gauss' law for a higher rank Coulomb liquid \cite{Pretko2017}.

It is thus apparent that the merging of topological defects
in the BZ generates a higher rank spin
liquid.
The existence of this higher rank 
liquid is confirmed by our Monte Carlo simulations 
of $S({\bf q}$ [Fig. \ref{fig:honeycomb_sq}], which 
exhibit four-fold pinch points at $\gamma=1/2$ 
\cite{Prem2018}.

However, this higher rank spin liquid
is unstable against varying $\gamma$.
For $\gamma>1/2$ the topological defects separate once again, and the four fold pinch point splits up into four
two-fold pinch points. 
Two of these per BZ remain at the BZ corners, while the others
migrate into the interior of the BZ.
The  process of appearance, merging and separation of the topological defects is similar to the behavior of Dirac points in models of graphene with further neighbor hopping \cite{bena11, montambaux12}.

In summary, the honeycomb model [Eq. (\ref{eq:Hhex})] exhibits three distinct classical spin liquids with zero
temperature transitions between them.
These are associated in one case to topological defect creation/annihilation and in the other
to a coalescence of defects leading to a higher-rank spin liquid.

\emph{Octochlore model}. We now turn to our second example:
a model on the three dimensional octochlore lattice [Fig.
\ref{fig:octochlore}].
The Hamiltonian is
\begin{eqnarray}
&&\mathcal{H}_{\sf o}=
\frac{J}{2} \sum_{{\rm oct}} {\bf M}_{{\rm oct}, \alpha, \beta}^2
\label{eq:Hoct}
\\
&&{\bf M}_{{\rm oct}, \alpha, \beta}=\sum_{i \in {\rm oct}} {\bf S}_i + \alpha \sum_{i \in \langle{\rm oct}\rangle} {\bf S}_i 
+ \beta \sum_{i \in \langle\langle{\rm oct}\rangle\rangle} {\bf S}_i 
\label{eq:Moct}.
\end{eqnarray}
The sum in Eq. (\ref{eq:Hoct}) is over octahedra.
The first sum in  Eq. (\ref{eq:Moct}) is over spins on the octahedron (red sites in Fig. \ref{fig:octochlore}(b)),   the second is over spins nearest to the outside of the octahedron (blue in Fig. \ref{fig:octochlore}(b)) and
the third over the next nearest spins to the octahedron
(yellow in Fig. \ref{fig:octochlore}(b)). 
$\alpha$ and $\beta$ are tuning parameters. Ground states obey the constraint ${\bf M}_{{\rm oct}, \alpha, \beta}=0$ everywhere.

\begin{table}[]
    \centering
    \begin{tabular}{c|c|c|c|c|}
         &   $(\pi, \pi, \pi)$ & $(\pi, \pi, \pi)+$& $(\pi, \pi, \pi)+$ & $(\pi, \pi, \pi)+$ \\
         &    & $(q,q,q)$& $(q,q,0)$ & $(q,0,0)$ \\
         \hline
    I$^{\pm}$     &  $\pm 1$ & 0 & 0 & 0 \\
    II$^{\pm}$     &  $\mp 1$ & $8 \times \pm 1$  & 0 & 0 \\
    III$^{\pm}$     &  $\mp 1$ & 0 & 0  & $6 \times \pm 1$ \\
    IV$^{\pm}$     &  $\mp 1$ & $8 \times \pm 1$ & $12 \times \mp 1$  & $6 \times \pm 1$ \\
    V$^{\pm}$     &  $\pm 1$ & $8 \times \mp1$ & $12 \times \pm 1$  & 0 \\
    VI$^{\pm}$     &  $\pm 1$ & 0 & $12 \times \pm 1$  & $6 \times \mp 1$ \\
    \end{tabular}
    \caption{Different octochlore CSLs on the 
    phase diagram of Fig.  \ref{fig:octochlore}(c),
    distinguished by the number and arrangement of
    topological defects in the Brillouin Zone. 
   Defects always appear at the Brillouin zone corners $(\pi,\pi,\pi)$ or positions displaced from the zone corner along high symmetry directions ${\bf q}$-space. Entries in
   the table refer to the number and charge of such defects in each CSL.
    }
    \label{tab:octo_table}
\end{table}

There are three sites, and one constraint,
per unit cell so there is one three-component constraint
vector ${\bf L}({\bf q})$. 
Inversion symmetry forces ${\bf L}({\bf q})$
to be real.
In three dimensions, this allows topological defects in  ${\bf L}({\bf q})$, with integer topological charge given by the skyrmion winding number of the normalized
vector $\tilde{{\bf L}}({\bf q})$ around closed 2D surfaces.

Varying $\alpha, \beta$ enables the generation of several liquids, distinguished through  number and arrangement of topological defects in reciprocal space.
The phase diagram is shown in Figure \ref{fig:octochlore}(c), with the definition of the different liquids in terms of the arrangement of defects in the BZ in Table \ref{tab:octo_table}.
Defects always appear at the zone corners
${\bf q}=(\pm\pi, \pm\pi, \pm\pi)$, and can
additionally appear at momenta displaced from
the zone corner along high symmetry directions.

Once again, defects in ${\bf L}({\bf q})$ correspond to
singularities in  $S({\bf q})$.
As seen in Table \ref{tab:octo_table}, some liquids feature many defects in the BZ, including up to 27
in the case of spin liquid IV.
This results in an abundance of pinch points in $S({\bf q})$ as shown in Fig. \ref{fig:octochlore_sq}(a).

Multifold pinch points indicating higher rank spin 
liquids emerge at certain boundaries, for example
the boundary between Liquids II and VI. In this case, starting from II, 8 $Q=-1$ defects placed
at $(\pi\pm \delta,\pi\pm \delta,\pi\pm \delta)$  converge on a $Q=1$ defect at ${\bf q}=(\pi,\pi,\pi)$.
The resulting $Q=-7$ defect at the transition results in a  complicated structure in $S({\bf q})$, which has the appearance of a 6-fold pinch point when cut through the $(h,h,l)$ plane
[Fig. \ref{fig:octochlore_sq}(b)].

\emph{Summary and Outlook}-We have demonstrated a new 
approach to the discovery of models exhibiting exotic
Coulomb liquids, including higher rank Coulomb liquids.
This approach is built on relating the ground state
constraints which define classical spin liquids to the
topological properties of a vector in reciprocal space.
We discover several new Coulomb liquids, all corresponding to simple Hamiltonians on local clusters, some with large
numbers of pinch points per Brillouin Zone in $S({\bf q})$,
and topological transitions between them, some of which exhibit
higher rank spin liquids connected to the potential 
realization of fractons.

The ability to distinguish classical spin liquids
using topological properties of the constraint 
vector ${\bf L}({\bf q})$ suggests the possibility of a 
comprehensive classification of classical spin liquids, in the spirit of the established classification of topological insulators and Weyl semi-metals
\cite{Schnyder2008, Ryu2010, Fu2011, Chiu2016}.
We will explore this classification in forthcoming work.

{\it Acknowledgements:}
O. B. thanks Han Yan for useful discussions.
This work was in part supported by the Deutsche Forschungsgemeinschaft  under grants SFB 1143 (project-id 247310070) and the cluster of excellence ct.qmat (EXC 2147, project-id 390858490).

\bibliography{references.bib}

\end{document}


\title{Supplemental Material: Topological Route to New and Unusual Coulomb Spin Liquids}

\author{Owen Benton}
\affiliation{Max Planck Institute for the Physics of Complex Systems, N{\"o}thnitzer Str. 38, Dresden 01187, Germany}

\author{Roderich Moessner}
\affiliation{Max Planck Institute for the Physics of Complex Systems, N{\"o}thnitzer Str. 38, Dresden 01187, Germany}

\maketitle

\section{Details of Monte Carlo simulations}

Here we give some details about the Monte Carlo simulation used to calculate $S({\bf q})$ 
for the honeycomb and octochlore models in Fig. 3 and 4 of the main text.

We simulate $O(3)$ vector spins with unit length and Hamiltonians given by Eq. (7) (honeycomb model) and Eq. (14)
(octochlore model) of the main text. The simulations use the Metropolis algorithm. One Monte Carlo step (MCS)
consists of one sweep of the lattice attempting a random reorientation of each spin in succession, accepting the update with probability 1 if it reduces the energy and accepting with probability $e^{-\delta E/T}$ if it increases the energy by $\delta E$.

For simulating $S({\bf q})$ we take a  cluster of $N$ sites, and seek to calculate $S({\bf q})$ at a temperature $T_0$.
The simulations are started at a higher temperature $T=T_{start}$ and gradually cooled down to the base temperature with $N_{eq}$ MCS at each of $N_T$ logarithmically spaced intermediate temperatures.
After a further $N_{eq}$ MCS for equilibriation at $T=T_0$,   $S({\bf q})$ is averaged over $N_{measure}$ measurements, each separated by $N_{sep}$ MCS.
The same procedure occurs over $N_{par}$ parallel runs of the simulation, with the final result reported being the average
of those runs.

For the honeycomb simulations we calculated $S({\bf q})$ at $T_0=0.002J$ on a cluster with $N=1920$ spins. The other settings of the simulation
were $T_{start}=10J$, $N_T=60, N_{eq}=10000, N_{sep}=1000, N_{measure}=6000, N_{par}=25$.

For the octochlore simulations we calculated $S({\bf q})$ at $T_0=0.01J$ on a cubic cluster with $N=31 944$ spins. The other settings of the simulation
were $T_{start}=10J$, $N_T=50, N_{eq}=5000, N_{sep}=2000, N_{measure}=1250, N_{par}=200$.

\section{Long wavelength theory of the honeycomb model at $\gamma=1/2$}

Here we give a derivation of Eq. (13) of the main text,
in which constraints on the honeycomb 
spin liquid at $\gamma=1/2$
are, after coarse graining, expressed as a constraint on the second derivatives of a traceless tensor field $m_{\mu \nu}$.
\begin{eqnarray}
\partial_{\mu} \partial_{\nu} m_{\mu \nu}=0
\end{eqnarray}

\begin{figure}
    \centering
    \includegraphics[width=0.4\textwidth]{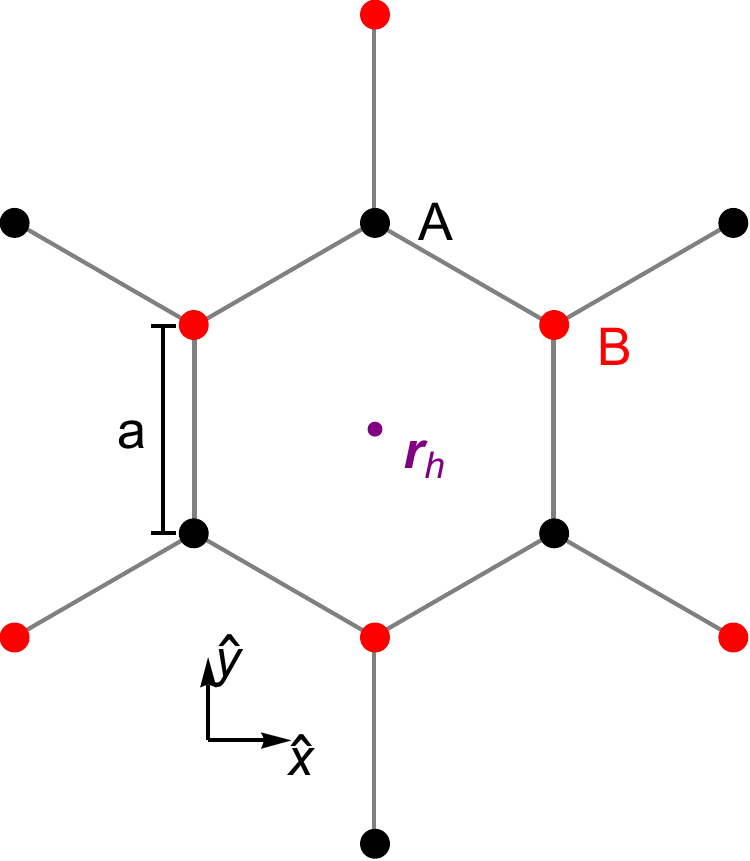}
    \caption{A hexagon and its exterior spins on the honeycomb lattice, with A and B sublattices indicated.}
   \label{fig:honeycomb_supp}
\end{figure}

To obtain this result, we consider the constraint on
a single hexagon $h$, centred on position ${\bf r}_{h}$
of the lattice [Fig. \ref{fig:honeycomb_supp}].

The ground state constraint on $h$ is
\begin{eqnarray}
\sum_{j \in h} S_j^{\sigma}+\gamma \sum_{j \in \langle h \rangle}S_j^{\sigma}=0 \quad \forall \sigma
\end{eqnarray}
where the first sum runs over spins belonging to $h$
and the second sum runs over spins connected to the exterior of $h$.
$\sigma$ is a spin component index. 
Since different spin components are not coupled by the constraint we drop this index from now on.
The only difference this makes is that for $O(\mathcal{N})$ spins we should remember that there are in principle $\mathcal{N}$ copies of each of the fields defined below, one for each spin component.

We introduce some complex variables $S_{K,j}$, defined
as 
\begin{eqnarray}
S_{K, j}=e^{i \mathbf{K} \cdot {\bf r}_j} S_j
\end{eqnarray}
where ${\bf r}_j$ is the position of site $i$ and
\begin{eqnarray}
{\bf K}= \left( \frac{4\pi}{3\sqrt{3} a}, 0\right)
\end{eqnarray}
with $a$ the honeycomb bond length.
We then further define
\begin{eqnarray}
S_{K, j}=t_j -i \zeta_j v_j
\end{eqnarray}
where $\zeta_j=\pm1$ is a sign factor which is +1 
for sites on the `A' sublattice and -1 for on the `B' sublattice (see Fig. \ref{fig:honeycomb_supp}),
and $t_j, v_j$ are real.

In terms of $t_j$ and $v_j$ the constraint is then
\begin{eqnarray}
\sum_{j \in h}
e^{-i \mathbf{K} \cdot {\bf r}_j} (t_j - i \eta_j v_j)
+\gamma \sum_{j \in \langle h \rangle}
e^{-i \mathbf{K} \cdot {\bf r}_j} (t_j - i \eta_j v_j)
=0 
\label{eq:tv-constraint}
\end{eqnarray}

To coarse grain, we now introduce continuous fields
$\tilde{t}({\bf r})$, $\tilde{v}({\bf r})$, defined
such that when they are evaluated at the lattice
sites ${\bf r}={\bf r}_j$, they return the values
of the $t_j$, $v_j$ on those sites.
Assuming smooth variation of $\tilde{t}({\bf r}),
\tilde{v}({\bf r})$, we can Taylor expand around 
the center of the hexagon ${\bf r}={\bf r}_h$
to obtain
\begin{eqnarray}
&&t_j\approx\tilde{t}({\bf r}={\bf r_h})+
\sum_{\mu} d_{\mu} \partial_{\mu} \tilde{t}|_{({\bf r}
={\bf r}_h)}
+
\sum_{\mu \nu} (r_j^{\mu}-r_h^{\mu})
(r_j^{\nu}-r_h^{\nu})\partial_{\mu}
\partial_{\nu} \tilde{t}|_{({\bf r}
={\bf r}_h)}\\
&&v_j \approx
\tilde{t}({\bf r}={\bf r_h})+
\sum_{\mu} d_{\mu} \partial_{\mu} \tilde{v}|_{({\bf r}
={\bf r}_h)}
+
\sum_{\mu \nu} (r_j^{\mu}-r_h^{\mu})
(r_j^{\mu}-r_h^{\nu}) 
\partial_{\mu}
\partial_{\nu} \tilde{v}|_{({\bf r}
={\bf r}_h)}
\end{eqnarray}

Inserting this into Eq. (\ref{eq:tv-constraint})
we obtain the coarse grained constraint
on $\tilde{t}({\bf r}), \tilde{v}({\bf r})$
\begin{eqnarray}
3 a i (2 \gamma -1) (\partial_x t + \partial_y v)
+ \frac{3 a^2 }{4} (1+4\gamma) 
(-\partial_x^2 t + \partial_y^2 t + 2 \partial_x \partial_y v)=0.
\end{eqnarray}

When $\gamma=1/2$, the first derivative term vanishes,
and we are left with
\begin{eqnarray}
(-\partial_x^2 t + \partial_y^2 t + 2 \partial_x \partial_y v)=0.
\label{eq:r2constraint}
\end{eqnarray}
If we define a traceless matrix:
\begin{eqnarray}
m=
\begin{pmatrix}
-t & v \\
v & t
\end{pmatrix}
\end{eqnarray}
Eq. (\ref{eq:r2constraint}) then takes exactly the 
form of Eq. (13) of the main text.
